\documentclass[a4paper,11pt]{article}

\pdfoutput=1
\usepackage{jcappub}

\usepackage{graphicx}
\usepackage{longtable}
\usepackage{float}
\usepackage{dcolumn}
\usepackage{bm}
\usepackage{appendix}
\usepackage{multirow}
\usepackage{color}
\usepackage[utf8]{inputenc}
\usepackage{footmisc}
\usepackage{hyperref}

\usepackage{xcolor}
\usepackage{graphicx}
\usepackage{bm}
\usepackage{ulem}
\usepackage{multirow, array, float, tabularx}
\usepackage{amsmath}
\hypersetup{
	citecolor = blue,
	linkcolor = blue,
	urlcolor = magenta,
}

\newcommand{\be}{\begin{equation}}
	\newcommand{\ee}{\end{equation}}
\newcommand{\bea}{\begin{eqnarray}}
	\newcommand{\eea}{\end{eqnarray}}
\newcommand{\bes}{\begin{subequations}}
	\newcommand{\ees}{\end{subequations}}

\newcommand{\bc}{\begin{center}}
	\newcommand{\ec}{\end{center}}

\usepackage{wasysym}

\begin{document}

\title{A New Window on Dynamical Dark Energy: Combining DESI-DR2 BAO with future Gravitational Wave Observations}

\author[a]{Felipe Bruno Medeiros dos Santos,}
\author[a]{Jonathan Morais,}
\author[b,c]{Supriya Pan,}
\author[d]{Weiqiang Yang,}
\author[e]{Eleonora Di Valentino}
\affiliation[a]{Observatório Nacional, Rua General José Cristino, 20921-400, Rio de Janeiro-RJ, Brazil}
\affiliation[b]{Department of Mathematics, Presidency University, 86/1 College Street, Kolkata 700073, India}
\affiliation[c]{Institute of Systems Science, Durban University of Technology, Durban 4000, Republic of South Africa}
\affiliation[d]{Department of Physics, Liaoning Normal University, Dalian, 116029, People's Republic of China}
\affiliation[e]{School of Mathematical and Physical Sciences, University of Sheffield, Hounsfield Road, Sheffield S3 7RH, United Kingdom}

\emailAdd{fbmsantos@on.br, jmorais@on.br, supriya.maths@presiuniv.ac.in, d11102004@163.com, e.divalentino@sheffield.ac.uk}

\abstract{Baryon acoustic oscillation (BAO) data from the Dark Energy Spectroscopic Instrument (DESI) appear to indicate the first evidence for dynamical dark energy (DDE), with a present-day behavior resembling quintessence. This evidence emerges when the Chevallier-Polarski-Linder (CPL) parameterization of the dark energy equation of state, $w_{\textrm{de}} = w_0 + w_a (1-a)$, is considered, and persists across other functional forms of $w_{\textrm{de}}$. In this work, we investigate how the inclusion of future gravitational wave (GW) standard siren data impacts the uncertainties in cosmological parameters when combined with DESI measurements. Specifically, we analyze the expected contributions from upcoming GW observatories such as the Einstein Telescope (ET) and the Deci-hertz Interferometer Gravitational-wave Observatory (DECIGO), as well as the current Laser Interferometer Gravitational-Wave Observatory (aLIGO). We find that the addition of GW data, particularly from DECIGO, significantly reduces the uncertainties in cosmological parameters, with the extent of the improvement depending on the specific form of $w_{\textrm{de}}$ and being more expressive for the $\Omega_m$ and $H_0$ parameters for all models studied. Our results highlight both the constraining power of future GW observations and the importance of considering a range of cosmological models in the data analysis.}

\maketitle

\section{Introduction}\label{sec:intro}

Dark Matter (DM) and Dark Energy (DE) are two main dark ingredients of our universe that occupy nearly 96\% of its total energy budget. The distributions of DM and DE, according to the latest astronomical datasets, are well described in the context of Einstein's General Relativity (GR) in terms of the $\Lambda$CDM cosmology, a model that has been extremely successful in explaining a series of astronomical and cosmological observations, such as the cosmic microwave background (CMB)~\cite{Planck:2018vyg}. On the other hand, it is also known that the model faces issues that suggest a departure from this standard picture~\cite{Weinberg:1988cp, Zlatev:1998tr}, the most relevant of which right now is the tension in the measurements of the present Hubble rate of expansion $H_0$~\cite{Verde:2019ivm,DiValentino:2020zio,DiValentino:2021izs,Perivolaropoulos:2021jda,Schoneberg:2021qvd,Shah:2021onj,Abdalla:2022yfr,DiValentino:2022fjm,Kamionkowski:2022pkx,Giare:2023xoc,Hu:2023jqc,Verde:2023lmm,DiValentino:2024yew,Perivolaropoulos:2024yxv}, between early- and late-time measurements~\cite{Freedman:2020dne,Birrer:2020tax,Wu:2021jyk,Anderson:2023aga,Scolnic:2023mrv,Jones:2022mvo,Anand:2021sum,Freedman:2021ahq,Uddin:2023iob,Huang:2023frr,Li:2024yoe,Pesce:2020xfe,Kourkchi:2020iyz,Schombert:2020pxm,Blakeslee:2021rqi,deJaeger:2022lit,Murakami:2023xuy,Breuval:2024lsv,Freedman:2024eph,Riess:2024vfa,Vogl:2024bum,Gao:2024kkx,Scolnic:2024hbh,Said:2024pwm,Boubel:2024cqw}. 
Perhaps the simplest approach to model new physics beyond the $\Lambda$CDM model is to introduce a dynamical DE (DDE) component, where the equation-of-state (EoS) of DE, $w_{\rm de}$, is either a constant ($\neq -1$), or varies with time $-$ widely known as the parametrized $w_{\rm de}$ models. Naturally, over the past years, a cluster of $w_{\rm de}$ models have been introduced in the literature~\cite{Cooray:1999da,Efstathiou:1999tm,Chevallier:2000qy,Astier:2000as,Weller:2001gf,Linder:2002et,Wetterich:2004pv,Hannestad:2004cb,SDSS:2004kqt,Jassal:2004ej,Alam:2004jy,Upadhye:2004hh,Gong:2005de,Linder:2005ne,Nesseris:2005ur,Liu:2008vy,Barboza:2008rh,Barboza:2009ks,Ma:2011nc,Sendra:2011pt,Li:2011dr,DeFelice:2012vd,Magana:2014voa,Rezaei:2017yyj,Yang:2017alx,Pan:2017zoh,Yang:2018prh,Mehrabi:2018oke,Yang:2018qmz,DiValentino:2019dzu,Li:2019yem,Pan:2019brc,Hernandez-Almada:2020uyr,Perkovic:2020mph,Yang:2021flj,Yang:2021eud,Alestas:2021luu,Colgain:2021pmf,Escamilla:2023oce,Najafi:2024qzm,Giare:2024ocw,Giare:2025pzu,Kessler:2025kju}. In most cases, $w_{\rm de}$ has been proposed on phenomenological grounds due to the unavailability of any fundamental principle that governs the parametrized $w_{\rm de}$ models. However, connections to more fundamental constructions can be achieved; for example, for each phenomenological $w_{\rm de}$ model, it is in principle possible to find an equivalent scalar field theoretic description, where the potential of the scalar field can be investigated~\cite{Scherrer:2015tra}. Also, one can link the behavior of $w_{\rm de}(a)$ to a corresponding interacting dark energy model, with essentially the same background and perturbative behavior~\cite{vonMarttens:2019ixw}.

Recently, results from the DESI (Dark Energy Spectroscopic Instrument) Year 1 survey~\cite{DESI:2024mwx} seem to point to a significant departure from $\Lambda$CDM for the first time. Specifically, when a DDE model is considered, evidence for a dynamical behavior of $w_{\rm de}(a)$ is shown at more than $2\sigma$ confidence level. This has sparked a strong debate about the validity of the results and possible implications for beyond $\Lambda$CDM models~\cite{Yin:2024hba,Chan-GyungPark:2024mlx,Ramadan:2024kmn,Pourojaghi:2024tmw,Giare:2024gpk,Giare:2024ocw,Jiang:2024xnu,RoyChoudhury:2024wri,Li:2024qus,Zheng:2024qzi,Giare:2025pzu,Wolf:2025jlc,Ormondroyd:2025exu,Ormondroyd:2025iaf,Pang:2025lvh,DESI:2025hce,Pan:2025psn,Ishiyama:2025bbd,Berti:2025phi,Nakagawa:2025ejs,Hur:2025lqc,Paliathanasis:2025dcr,Brandenberger:2025hof,Plaza:2025ryz,Nesseris:2025lke}. The trend has been confirmed in the newest release from the collaboration (DR2)~\cite{DESI:2025zgx}, in which a deviation from $\Lambda$CDM is seen at the $3.1\sigma$ level for CMB and DESI data only. The result holds even when using supernova (SNe) data, though the preference varies depending on the dataset. Parallel to this, developments in future experiments seek to increase the precision of cosmological observables. In particular, measurements of gravitational waves from merging binary systems of black holes and neutron stars~\cite{LIGOScientific:2017vwq,LIGOScientific:2017zic}, with a number of events that might surpass thousands, place the third-generation (3G) observatories~\cite{Zhao:2010sz,Sathyaprakash:2012jk,Maggiore:2019uih,Seto:2001qf,Kawamura:2006up,Kawamura:2011zz}, located both on the ground and in space, as some of the best candidates to characterize and distinguish cosmological scenarios.

In the present work, we explore the possible impact of Gravitational Wave Standard Sirens (GWSS) in light of the newest DESI results. Given that we focus on the improvement in parameter uncertainties, we take the following approach: we simulate GWSS taking each DDE model as the fiducial one; this allows us to estimate the impact on the uncertainties of cosmological parameters given the configuration from different observatories. We restrict our analysis to three well-known parametrized $w_{\rm de}$ models, namely the Chevallier-Polarski-Linder (CPL)~\cite{Chevallier:2000qy,Linder:2002et}, the Barboza-Alcaniz (BA)~\cite{Barboza:2008rh}, and the Jassal-Bagla-Padmanabhan (JBP)~\cite{Jassal:2005qc} models. As for the gravitational wave observatories, we generate mock datasets from the Einstein Telescope (ET), a future ground-based gravitational wave detector~\cite{Punturo:2010zz,Maggiore:2019uih}, the Deci-hertz Interferometer Gravitational wave Observatory (DECIGO), a space-based one~\cite{Seto:2001qf,Kawamura:2006up,Kawamura:2011zz}, and the Laser Interferometer Gravitational-Wave Observatory (aLIGO) collaboration, a currently operating ground-based detector~\cite{LIGOScientific:2017vwq,LIGOScientific:2017zic}. 

The paper is organized as follows. In Section~\ref{sec:2}, we briefly introduce the dynamical DE models. Sections~\ref{sec:3} and \ref{sec:4} describe the observational datasets and the methodology adopted in this work. In Section~\ref{sec:5}, we present the constraints on the models. Finally, in Section~\ref{sec:6}, we close the article with a discussion of our findings.

\section{Dynamical dark energy}\label{sec:2}

One of the most direct ways of modeling the behavior of the dark sector in the universe is through the parameterization of the respective component. This has been done extensively over the years for the dark energy fluid~\cite{Efstathiou:1999tm,Chevallier:2000qy,Astier:2000as,Weller:2001gf,Linder:2002et,Wetterich:2004pv,SDSS:2004kqt,Jassal:2004ej,Alam:2004jy,Upadhye:2004hh,Gong:2005de,Linder:2005ne,Barboza:2008rh,Barboza:2009ks,Ma:2011nc,Sendra:2011pt,Li:2011dr,DeFelice:2012vd,Rezaei:2017yyj,Yang:2017alx,Yang:2018prh,Mehrabi:2018oke,Yang:2018qmz,Pan:2019brc,Hernandez-Almada:2020uyr,Yang:2021flj,Yang:2021eud,Escamilla:2023oce}, for many different purposes and observational probes. By considering a homogeneous and isotropic spacetime characterized by the spatially flat Friedmann-Lema\^{i}tre-Robertson-Walker (FLRW) metric, one obtains the Friedmann equations as solutions of the Einstein field equations. The Hubble equation connecting the total energy density of the universe with its expansion follows as
\begin{equation}
    H^2 = \frac{8\pi G}{3} \left(\rho_\textrm{m,0} a^{-3} + \rho_\textrm{r,0} a^{-4} + \rho_{\nu} + \rho_{\textrm{de}} \right),
    \label{eqn:Hubble}
\end{equation}
with $\rho_\textrm{m,0}$ and $\rho_\textrm{r,0}$ being the current energy densities of matter and radiation, respectively, evolving as a function of the scale factor $a$ of the FLRW universe (we assume $a_0=1$). $H(a)=\dot a/a$ is the Hubble parameter in terms of the cosmic time $t$; $\rho_{\nu}$ is the energy density of the neutrino sector,\footnote{As commonly done in the literature, we fix the sum of neutrino masses to $\sum m_\nu = 0.06$~eV, while the number of neutrino species is fixed to $N_{\textrm{eff}} = 3.044$.} while $\rho_{\textrm{de}}$ can be determined as a solution of the continuity equation, given a parameterization for the equation-of-state parameter $w(a)$, such that the evolution of dark energy is generally given as
\begin{equation}\label{eqn:rhode}
\rho_{\textrm{de}} = \frac{\rho_{\textrm{de,0}}}{a^3} \exp\left(-3\int^a_1 \frac{w_{\textrm{de}}(a')}{a'} \, da'\right),
\end{equation}
where $\rho_{\rm de, 0}$ is the present value of the DE density, and the choice of a given form for $w(a)$ completes the requirements necessary to determine the cosmological evolution given by Eq.~(\ref{eqn:Hubble}). There is significant freedom in the choice of the function $w_{\rm de}(a)$, which can be used to approximate the behavior of several dark energy models, especially quintessential ones, in which the dark energy presence is explained by a scalar field active at late times~\cite{Tsujikawa:2013fta}.

Now we present the specific forms studied in this work. We focus on two-parameter models characterized by the following free parameters: $w_0$, which corresponds to the present value of $w_{\rm de}(a)$, and $w_a = -\frac{dw_{\rm de}(a)}{da}\big|_{a = a_0}$, which quantifies the dynamical (or non-dynamical) nature of $w_{\rm de}(a)$. The standard $\Lambda$CDM model can be recovered for the choices $w_0 = -1$ and $w_a = 0$:
\begin{itemize}
    \item \textbf{Chevallier-Polarski-Linder (CPL) model:} The CPL model~\cite{Chevallier:2000qy,Linder:2002et} is the most studied parameterization of dark energy, given by a Taylor expansion of $w_{\rm de}(a)$ around $a = a_0 = 1$ up to the first-order term. It was also the model adopted by the DESI team to report their first results on dynamical dark energy. The form for $w_{\rm de}(a)$ is:
    \begin{equation}
        w_{\rm de}(a) = w_0 + w_a(1 - a).
    \end{equation}
    
    \item \textbf{Barboza-Alcaniz (BA) model:} The BA model~\cite{Barboza:2008rh} addresses the issue of the CPL model diverging as $a \rightarrow \infty$, by providing a finite behavior. It also allows a phantom crossing of $w_{\rm de}(a)$ and satisfies $w_{\rm de}(a) = w_0 + w_a$ in the limit $a \rightarrow 0$:
    \begin{equation}
        w_{\rm de}(a) = w_0 + w_a \frac{1 - a}{a^2 + (1 - a)^2}.
    \end{equation}
    
    \item \textbf{Jassal-Bagla-Padmanabhan (JBP) model:} The JBP model~\cite{Jassal:2004ej}, with the same notations as described above, has the following form:
    \begin{equation}
        w_{\rm de}(a) = w_0 + w_a \, a(1 - a).
    \end{equation}
    Note that in this case, $w_{\rm de}(a)$ is a second-degree polynomial in $a$.
\end{itemize}

For all parameterizations, using Eqs.~\ref{eqn:Hubble} and~\ref{eqn:rhode}, one can, in principle, determine the expansion history of the universe at the background level. On the other hand, understanding the expansion history at the level of perturbations driven by these parametrized forms of $w_{\rm de}(a)$ is also important and necessary. We have followed the standard equations as described in Ref.~\cite{Ma:1995ey}. Thus, having both expansion histories at the background and perturbation levels, one can completely describe the behavior of the proposed DE parameterizations.

\section{Methodology and Datasets}\label{sec:3}

We use the dynamical DE implementation built into the \texttt{CLASS} code~\cite{Blas:2011rf,Lesgourgues:2011re}, while sampling the parameter space using \texttt{MontePython}~\cite{Brinckmann:2018cvx}. We vary the standard cosmological parameters of the $\Lambda$CDM model, along with the two DDE parameters that characterize each parameterization, $w_0$ and $w_a$. The uncertainties of all parameters and the confidence contour plots were obtained with the \texttt{GetDist} code~\cite{Lewis:2019xzd}, which analyzes the resulting chains.

For the cosmological data, we use the following: CMB data from the Planck collaboration~\cite{Planck:2019nip}; specifically, we combine the low-multipole TT and EE modes together with higher-multipole data ($\ell > 30$), which includes the correlated TE modes, represented by the \texttt{plik} likelihood. We also include the reconstructed lensing potential obtained from the 3-point correlation function of the Planck data~\cite{Planck:2018lbu}. These are combined with late-time measurements from the PantheonPlus catalog~\cite{Scolnic:2021amr,Brout:2022vxf}, composed of 1701 light curves from 1550 Type Ia supernovae (SNe).
Additionally, we consider the recent DESI BAO DR2 data~\cite{DESI:2025zgx}, in order to construct our baseline dataset (Base~$\equiv$~CMB+DESI+PantheonPlus). These consist of measurements of baryonic acoustic oscillations in the clustering of galaxies, quasars, and the Lyman-$\alpha$ forest at high redshifts, categorized into six different types of tracers. These measurements can be translated into geometrical quantities, as displayed in Table~4 of Ref.~\cite{DESI:2025zgx}. One of the most striking results regarding these data is the apparent preference for a dynamical dark energy component with respect to the $\Lambda$CDM model, especially when combined with different SNe samples.

Finally, in addition to these observables, we perform a joint analysis incorporating future expectations from GWSS~\cite{Holz:2005df}. The detection of a gravitational wave signal enables a highly accurate estimation of the corresponding luminosity distance $d_L$. The redshift $z$ can be determined if the binary system has an electromagnetic counterpart; in such cases, these events are referred to as bright sirens. A notable example is the event GW170817~\cite{LIGOScientific:2017vwq}, a Binary Neutron Star (BNS) merger that was accompanied by a short gamma-ray burst and a kilonova, allowing the redshift to be inferred from its host galaxy, NGC 4993. In this work, we focus on using this type of event, whose mock data set production will be described in more detail in the following section.

\section{Generating the GW mock data sets}\label{sec:4}

The first crucial step in the generation GWSS catalogues is to model the astrophysical distribution of sources as a function of redshift. In the case of BNS systems, the merger rate is obtained through the convolution of the star formation rate (SFR) with a time-delay distribution, typically taken as $P(t_d) \propto 1/t_d$, with a minimum delay of $\sim 20\,\mathrm{Myr}$~\cite{Regimbau:2012}. Following~\cite{Iacovelli:2022bbs}, the resulting distribution is accurately approximated by re-fitting the convolved rate to a Madau--Dickinson parametrization~\cite{Madau:2014bja}, yielding the parameters $\alpha_z = 1.42$, $\beta_z = 4.62$, and $z_p = 1.84$. This leads to the redshift evolution function
\begin{equation} \label{Pz_GWs}
    \psi(z) \propto \frac{(1+z)^{1.42}}{1+\left(\frac{1+z}{1+1.84}\right)^{6.04}},
\end{equation}
which provides an accurate description of the BNS merger rate evolution after accounting for delay-time effects.

The observed redshift distribution of merger events is then obtained by combining the intrinsic merger rate evolution with cosmological volume and time-dilation effects, leading to
\begin{equation}
    P(z) \propto \psi(z)\,\frac{1}{1+z}\,\frac{dV_c}{dz}(z),
\end{equation}
where $dV_c/dz$ is the comoving volume element, and the factor $(1+z)^{-1}$ accounts for the conversion between the source-frame time and the observer-frame time.

Using \eqref{Pz_GWs}, we generate mock catalogues by randomly sampling a fixed number $N$ of redshift values, depending on the specific experiment under consideration. Since our analysis is based on a predetermined number of detected events, the overall normalization of the merger rate, including the local rate $R_0$, is not required and is therefore omitted. For aLIGO, we adopt $N = 100$, following the findings of~\cite{Chen:2017rfc}, where it was shown that a sample of 100 BNS events with detected electromagnetic counterparts can provide cosmology-independent constraints on $H_0$ with an accuracy of approximately $3\%$. Although the exact number of bright sirens expected in future observing runs of aLIGO remains uncertain, this threshold serves as a reasonable estimate. It is important to note that the use of aLIGO in this context is solely for comparative purposes, as it does not classify as a third-generation GW experiment, and we do not expect a significant improvement in the current constraints, given a more realistic number of events.

For the ET interferometer, the number $N$ is selected based on the expected number of detections within a given observation time window. It is anticipated that approximately 1000 bright siren events will be detected over a 3-year observation period~\cite{Belgacem:2018lbp}. As for DECIGO, there is still no consensus in the literature regarding the expected number of bright siren events. Therefore, we adopt the same threshold used for the ET, as this number is consistent with the expectations for the next generation of interferometers~\cite{Sathyaprakash:2009xs, Cai:2017aea}.

\begin{figure*}
\centering
\includegraphics[width=0.6\textwidth]{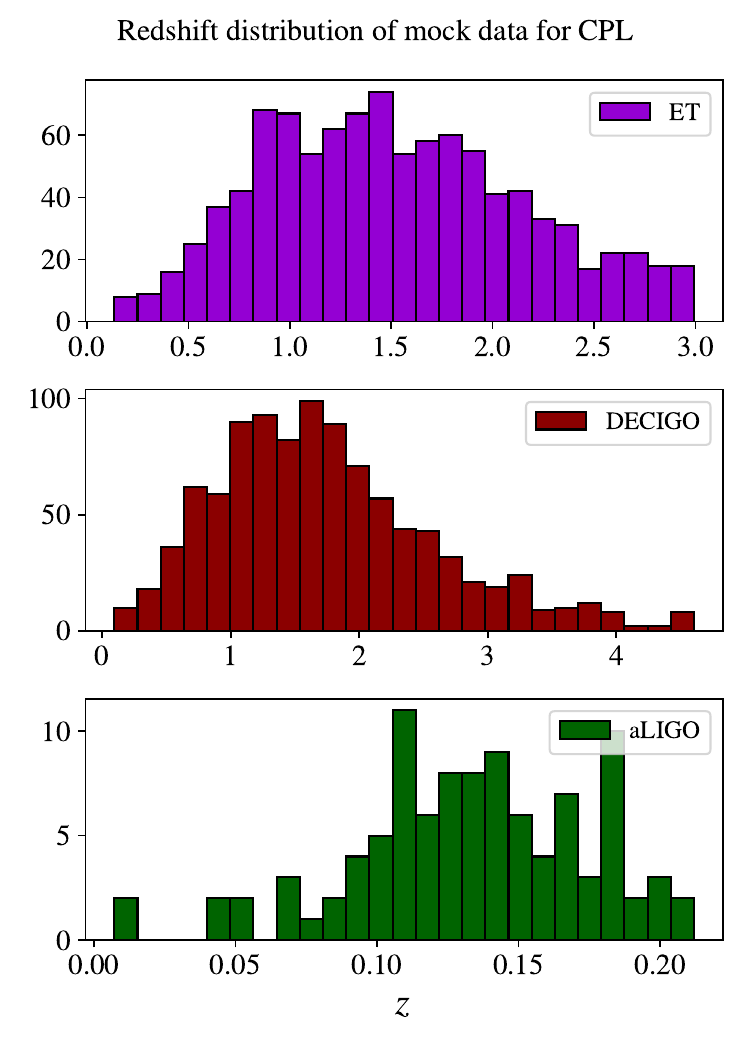}
\caption{Redshift distribution of the mock GW measurements according to the number of events, assuming a fiducial CPL model. We show the results for ET (purple), DECIGO (red), aLIGO (green).}
\label{fig:1}
\end{figure*}

The redshift detection limit for each interferometer is determined by its instrumental sensitivity. For the configurations adopted in the analysis, the detection horizon of aLIGO reaches $z \sim 0.2$, in agreement with previous studies~\cite{Lagos:2019kds}. In contrast, the Einstein Telescope extends its detection capability up to $z \sim 3$, consistent with the sensitivity projections reported in the ET design report updates~\footnote{\url{https://apps.et-gw.eu/tds/?r=18715}}. Owing to its superior sensitivity, DECIGO achieves a detection limit of approximately $z \sim 5$, in agreement with theoretical expectations~\cite{Zheng:2020tau}. The redshift distribution for each experiment, considering the CPL cosmological model, is shown in Fig.~\ref{fig:1}.

Given the sampled redshift values, the luminosity distance for each point is computed. It is defined as

\begin{equation}
    d_L(z) = (1+z) \, c \int_0^z \frac{dz'}{H(z')} \, ,
\end{equation}
where $c$ is the speed of light and $H(z)$ is the fiducial cosmology for the simulations.

While the fiducial cosmology is often chosen to be the $\Lambda$CDM model, as adopted in many works~\cite{Zhao:2010sz,Wang:2018lun,Du:2018tia,Matos:2021qne,Pan:2021tpk,Rezaei:2023xkj,Liu:2024bre,Liu:2024gne}, in this study, as previously described, we consider each DDE model as the fiducial one, given the evidence for dark energy dynamics reported by DESI. The fiducial parameters $\Omega_m$, $H_0$, $w_0$, and $w_a$ for each model under consideration can be found in Table~\ref{tab:1}.

\begin{table*}[t]
    \centering
    \renewcommand{\arraystretch}{1.3}
    \begin{tabular}{cccc}
        \hline
        \hline
        $\Omega_m$ & $H_0$ [km s$^{-1}$ Mpc$^{-1}$] & $w_0$ & $w_a$ \\
        \hline
        \hline
        \multicolumn{4}{c}{\textbf{CPL}} \\
        \hline
        $0.3096$ & $67.61$ & $-0.838$ & $-0.59$ \\
        \hline
        \multicolumn{4}{c}{\textbf{BA}} \\
        \hline
        $0.3092$ & $67.64$ & $-0.863$ & $-0.283$ \\
        \hline
        \multicolumn{4}{c}{\textbf{JBP}} \\
        \hline
        $0.3080$ & $67.65$ & $-0.804$ & $-1.13$ \\
        \hline
        \hline
    \end{tabular}
    \caption{Fiducial parameters used to generate the mock GW data for each DDE model.}
    \label{tab:1}
\end{table*}

With all the relevant information about the gravitational wave sources at hand, the next step is to simulate how these signals would be detected by the equipment. In the transverse-traceless (TT) gauge, the detector's response to a GW signal is expressed as
\begin{equation}
    h(t) = F_{+}(\theta,\phi,\psi)\,h_{+}(t) + F_{\times}(\theta,\phi,\psi)\,h_{\times}(t),
\end{equation}
where $F_{+}$ and $F_{\times}$ denote the antenna pattern functions corresponding to the two GW polarizations; the corresponding expressions for each experiment are shown in Appendix \ref{appendix:A}. The polarization components are given by $h_{+} = h_{xx} = -h_{yy}$ and $h_{\times} = h_{xy} = h_{yx}$. The angles $(\theta, \phi)$ define the direction to the source in the detector’s reference frame, while $\psi$ represents the polarization angle. For further details, see~\cite{Sathyaprakash:2009xs}. It is important to note that the form of the pattern functions depends on the orientation and the opening angle between the arms of the interferometer.

Next we compute the Fourier transform of the gravitational wave signal by applying the stationary phase approximation,
\begin{equation} \label{phase_eq}
    \mathcal{H}(f) = \mathcal{A}f^{-7/6}e^{i\Psi(f)},
\end{equation}
where the amplitude $\mathcal{A}$ is given by
\begin{equation} \label{amplitude}
\mathcal{A} = \frac{1}{d_L} \sqrt{F_+^2 (1 + \cos^2(\iota))^2 + 4 F_\times^2 \cos^2(\iota)} \, \sqrt{\frac{5\pi}{96\pi}} \, \mathcal{M}_c^{5/6} \, \pi^{-7/6},
\end{equation}
where $\iota$ denotes the inclination angle between the orbital plane and the observer's line of sight~\cite{Zhang:2019loq}. The quantity $\mathcal{M}_c$ corresponds to the observed chirp mass, defined as $\mathcal{M}_c = (1+z)M\eta^{3/5}$, where $M = m_1 + m_2$ is the total mass of the binary system, and $\eta = m_1 m_2 / M^2$ denotes the symmetric mass ratio. The phase in Eq.~\eqref{phase_eq} is computed within the post-Newtonian formalism up to 3.5 PN order, and the explicit expression can be found in~\cite{Nishizawa:2010xx}. Therefore, spin effects can be neglected when modeling the binary system.

Now that the amplitude detected by an interferometer can be calculated, it is necessary to verify whether the observed signal corresponds to a genuine gravitational wave or is merely instrumental noise. A GW detection is claimed only when the signal-to-noise ratio (SNR) of the detector network exceeds a threshold of 8, consistent with the current criterion adopted by the LIGO/VIRGO collaborations ~\cite{Cai:2017aea}.

For a network of $N$ independent interferometers, the combined SNR is given by
\begin{equation} \label{SNR}
    \rho = \sqrt{\sum_{i=1}^{N} \left( \rho^{(i)} \right)^2},
\end{equation}
where $\rho^{(i)} = \sqrt{\langle \mathcal{H}^{(i)}, \mathcal{H}^{(i)} \rangle}$ represents the individual SNR in the $i$-th detector~\cite{Sathyaprakash:2009xs}.

To compute the inner product in Eq.~\eqref{SNR}, we adopt the following prescription. Given $\tilde{a}(f)$ and $\tilde{b}(f)$ as the Fourier transforms of two time-domain functions $a(t)$ and $b(t)$, the scalar product is defined as
\begin{equation}
\langle a, b \rangle \equiv 4 \int_{f_{\text{min}}}^{f_{\text{max}}} \frac{\tilde{a}(f)\tilde{b}^*(f) + \tilde{a}^*(f)\tilde{b}(f)}{2} \frac{df}{S_h(f)},
\end{equation}
where $S_h(f)$ denotes the one-sided noise power spectral density (PSD). The integration limits \( f_{\min} \) and \( f_{\max} \) depend on the detector under consideration. Updated PSD curves for each detector can be found in the public documentation of the equipment: for aLIGO,\footnote{\url{https://dcc.ligo.org/LIGO-P1600143/public}} the ET\footnote{\url{https://www.et-gw.eu/index.php/etsensitivities}} (specifically the D sensitivity~\cite{Hild:2010id}), and DECIGO as reported by~\cite{Yagi:2009zz}.

Once the presence of the signal is confirmed through Eq.~\eqref{SNR}, the next and final step is to estimate
the uncertainty in the measurement of the luminosity distance.
We start by considering the instrumental uncertainty on $d_L$, which is evaluated using the standard
Fisher information matrix formalism~\cite{Tegmark:1996bz}.

The Fisher matrix $F_{ij}$ is defined as
\begin{equation}
F_{ij} =
\left\langle
\frac{\partial \mathcal{H}}{\partial \theta_i}
\bigg|
\frac{\partial \mathcal{H}}{\partial \theta_j}
\right\rangle ,
\end{equation}
where $\theta$ denotes the full set of intrinsic and extrinsic
waveform parameters entering the strain signal $\mathcal{H}$.

In the high signal-to-noise ratio limit, the inverse of the Fisher matrix provides an estimate of the
covariance matrix of the parameters,
\begin{equation}
\Sigma_{ij} = \left(F^{-1}\right)_{ij},
\end{equation}
with the uncertainty associated with a given parameter obtained from the corresponding diagonal element of the covariance matrix.

In particular, the marginalized $1\sigma$ uncertainty on the luminosity distance is given by
\begin{equation}
\frac{\Delta d_L^{(0)}}{d_L^{(0)}} = \Delta \theta_a
= \sqrt{\left(F^{-1}\right)_{aa}},
\end{equation}
where $\theta_a \equiv d_L^{(0)}$, and the superscript $(0)$ denotes the fiducial (true) value of the parameter. This expression corresponds to a fully marginalized error and therefore accounts for the covariances
between the luminosity distance and all other parameters included in the Fisher matrix, such as the
binary inclination, polarization angle, and sky location. All Fisher matrix calculations were performed using \texttt{gwfast},\footnote{\url{https://gwfast.readthedocs.io/en/latest/index.html}}
with further details on the numerical implementation described in
Refs.~\cite{Iacovelli:2022mbg, Iacovelli:2022bbs}.

Another relevant source of uncertainty is $\sigma^{\text{lens}}_{d_L}$, arising from weak lensing. A fitting formula is presented in~\cite{Tamanini:2016zlh}, based on~\cite{Hirata_2010}, and reads
\begin{equation}
    \sigma^{\text{lens}}_{d_L} = d_L (z) \times 0.066\left(\frac{1-(1+z)^{-0.25}}{0.25}\right)^{1.8}.
\end{equation}

Additionally, the contribution from peculiar velocities, associated with galaxy clustering and the binary barycentric motion, is also taken into account~\cite{Gordon:2007zw}, and is given by
\begin{equation}
    \sigma^{\text{pv}}_{d_L}(z) = d_L(z) \times \left| 1 - \frac{(1+z)^2}{H(z) d_L(z)} \right| \sigma_{v, \text{gal}},
\end{equation}
where $\sigma_{v, \text{gal}}$ is the one-dimensional velocity dispersion of the galaxy and is set to be $\sigma_{v, \text{gal}} = 300\, \mathrm{km\,s^{-1}}$, independent of the redshifts. Therefore, the total uncertainty on the measurement of $d_L$ is
\begin{equation}
    \sigma_{d_L} = \sqrt{(\sigma^{\text{inst}}_{d_L})^2 + (\sigma^{\text{lens}}_{d_L})^2 + (\sigma^{\text{pv}}_{d_L})^2}.
\end{equation}

After computing the values of $d_L$ and $\sigma_{d_L}$, the next natural step would be to perform a Monte Carlo simulation assuming a Gaussian distribution centered on these fiducial $d_L$ values, as done in previous works~\cite{Zhao:2010sz,Cai:2016sby,Wang:2018lun,Du:2018tia,Yang:2019vni,Yang:2019bpr,Lagos:2019kds,Baker:2020apq,Yang:2020wby,Geng:2020wns,Cao:2021jpx,Matos:2021qne,Pan:2021tpk,Rezaei:2023xkj}. However, to avoid discrepancies between the central values from GW data and EM data in the parameter planes, we omit this step, as done in Ref.~\cite{Zhang:2019loq}.

Having presented all the useful information related to the GWSS events, we now proceed to the final part dealing with the likelihood analysis. The likelihood function corresponding to the $N$ GWSS events is given by
\begin{equation}
    \ln\mathcal{L}_{\textrm{GW}} = -\frac{1}{2}\sum_{i=1}^N\left[\frac{d^{\textrm{obs}}_L(z_i)-d_L^{\textrm{th}}(z_i)}{\sigma_{d_L}}\right]^2,
\end{equation}
with $d^{\textrm{obs}}_L(z_i)$ and $d_L^{\textrm{th}}(z_i)$ corresponding to the mock data generated and the theoretical value for each model, respectively, for redshifts $z_i$. We combine the GWSS dataset with other cosmological probes to constrain the DE parameters.

\begin{table*}[t]
		\centering
		\begin{tabular}{>{\scriptsize}c >{\scriptsize}c>{\scriptsize}c>{\scriptsize}c>{\scriptsize}c}
			\hline
			\hline

            Dataset & $\sigma(\Omega_m)$ & $\sigma(H_0)$ & $\sigma(w_0)$ & $\sigma(w_a)$ \\
            \hline
            \hline 
            \multicolumn{5}{c}{\scriptsize{{CPL}}}\\
            \hline           
            Base & $0.0057$ & $0.57$ & $0.053$ & $^{+0.22}_{-0.18}$ \\
            Base+aLIGO & $0.0056$ & $0.57$ & $0.053$ & $^{+0.22}_{-0.19}$ \\
            Base+ET & $0.0051$ & $0.51$ & $0.054$ & $0.20$ \\
            Base+DECIGO & $0.0042$ & $0.38$ & $0.048$ & $0.18$ \\

            \hline
            \multicolumn{5}{c}{\scriptsize{{BA}}}\\ 
            \hline 
            Base & $^{+0.0052}_{-0.0063}$ & $^{+0.65}_{-0.58}$ & $0.048$ & $0.1$ \\
            Base+aLIGO & $0.0056$ & $0.58$ & $0.048$ & $0.1$ \\
            Base+ET & $0.0053$ & $0.55$ & $0.046$ & $^{+0.10}_{-0.093}$ \\
            Base+DECIGO & $0.0044$ & $0.42$ & $0.042$ & $^{+0.089}_{-0.08}$ \\

            \hline
            \multicolumn{5}{c}{\scriptsize{{JBP}}}\\ \hline 
            Base & $0.0057$ & $0.6$ & $0.077$ & $0.46$ \\
            Base+aLIGO & $0.0056$ & $0.58$ & $0.077$ & $0.46$ \\
            Base+ET & $0.0054$ & $0.56$ & $0.075$ & $0.44$ \\
            Base+DECIGO & $0.0045$ & $0.43$ & $0.07$ & $0.4$ \\

            \hline
            \hline
		\end{tabular}
		\caption{The 68\% confidence level (C.L.) uncertainties for the cosmological parameters obtained from the different data sets analyzed in this work. Note that Base refers to the combination of CMB+DESI+PantheonPlus data sets. }
        \label{tab:2}
\end{table*}
\begin{figure*}
    \centering
    \includegraphics[width=0.45\linewidth]{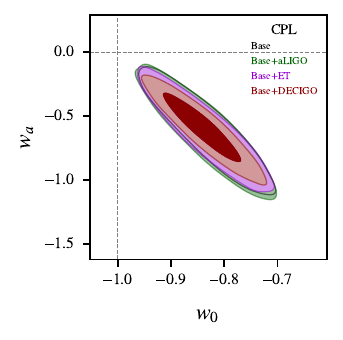}
    \includegraphics[width=0.45\linewidth]{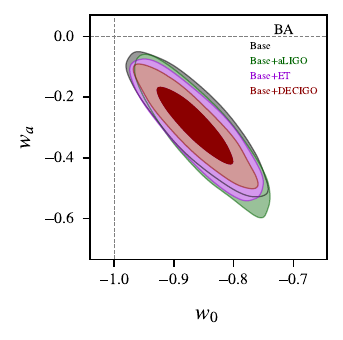} 
    \includegraphics[width=0.45\linewidth]{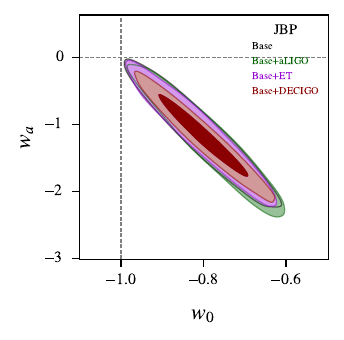} 
    \caption{The $w_0$–$w_a$ 68\% and 95\% confidence level contours for the CPL model (upper left panel), BA model (upper right panel), and JBP model (lower panel). We show results for the Base dataset (CMB+DESI+PantheonPlus) in gray, and its combinations with future GWSS experiments: Base+aLIGO (green), Base+ET (purple), Base+DECIGO (red).}
    \label{fig:2}
\end{figure*}

\begin{figure*}
    \centering
    \includegraphics[width=0.45\linewidth]{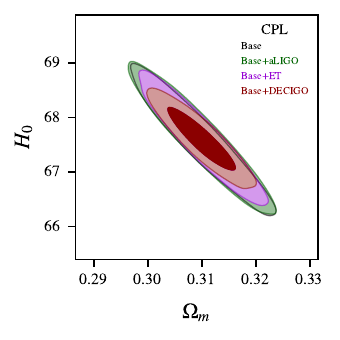}
    \includegraphics[width=0.45\linewidth]{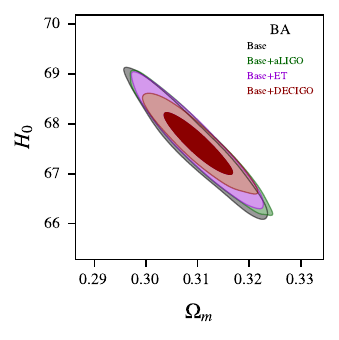} 
    \includegraphics[width=0.45\linewidth]{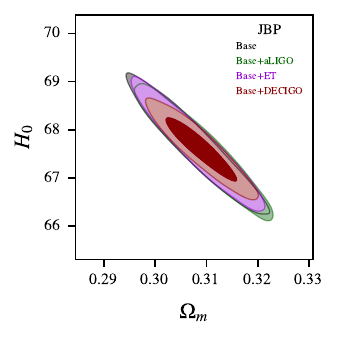} 
    \caption{The $\Omega_m$–$H_0$ 68\% and 95\% confidence level contours for the CPL model (upper left panel), BA model (upper right panel), and JBP model (lower panel), where $H_0$ is shown in units of km/s/Mpc. We show results for the Base dataset (CMB+DESI+PantheonPlus) in gray, and its combinations with future GWSS experiments: Base+aLIGO (green), Base+ET (purple) and Base+DECIGO (red).}
    \label{fig:3}
\end{figure*}

\section{Improvement in the cosmological parameters from DESI+GWSS}
\label{sec:5}

We have then performed analyses for our baseline data combination together with projections from ET, DECIGO, and aLIGO. The results are shown in Table~\ref{tab:2} and Figs.~\ref{fig:2} and~\ref{fig:3}. We focus on the improvements in the parameter uncertainties when each GW dataset is considered. Table~\ref{tab:2} presents the uncertainties at the $1\sigma$ confidence level (C.L.) for each $w_{\rm de}(a)$ parameterization.

We start with the CPL model. We note that the error reduction slightly improves progressively in the $w_0-w_a$ plane for Base+ET, and Base+DECIGO, respectively, with all parameters being affected at some level. We do not expect that a current-stage detector, such as aLIGO, will improve the constraints at a significant level, and that is perceived in our results for the DDE parameters $w_0$ and $w_a$, which basically stay at the same level of constraint as the current baseline data. We attribute this to the lower number of events considered, as well as the low redshift range, of around $z \leq 0.2$, which could in principle, result in better estimates for $H_0$ or $\Omega_m$, however, we do not see such behavior in our analysis. On the other hand, ET data span a wider redshift range and have more points, but with fewer points at lower redshifts, and also predict the largest errors for $d_L$ of all samples considered. These factors affect the constraints in a manner such that the $w_{de}$ parameter uncertainties are just slightly improved when compared to the analysis with aLIGO. In contrast, we also expect that DECIGO leads to the best constraints, given the sample size and decreased errors in the predicted detections. In fact, uncertainties for the matter fraction $\Omega_m$ decrease by around $26\%$, and the improvement in the $H_0$ estimate is quite impressive, with errors decreasing from $\sigma(H_0) = 0.57$ to $\sigma(H_0) = 0.38$, comparable with some future CMB observation forecasts. The $w_0$–$w_a$ plane is also better restricted, although not significantly, especially for the $w_0$ parameter, indicating the importance of GWSS data in probing further deviations from the standard model, as $\sigma(w_0)$ goes from $0.053$ to $0.048$, for instance\footnote{It is interesting to note that previous studies show that experiments such as Euclid can already place similar constraints on the CPL model at the same level as our Base+DECIGO analysis, for example~\cite{Euclid:2021qvm}.}.

For the BA model, the results are quite similar. Among the three forms of $w_{\rm de}(a)$ considered, this one yields the tightest constraints on $w_0$ and $w_a$, suggesting that it should also benefit the most from the inclusion of GWSS data. We see, however, that this is the case only for certain interferometers. When the constraints from the Base dataset for the $w_0$ and $w_a$ parameters are compared to those obtained with the one combined with ET, the improvement is mild, being more evident for the $w_0$ parameter, as seen in Table~\ref{tab:2}. However, in the same way as for the CPL model, constraints on $\Omega_m$ and $H_0$ are improved. Similarly as in the CPL case, the Base+aLIGO combination gives essentially the same results for the $w_0$–$w_a$ plane, although there is a visible improvement for the $\Omega_m$ and $H_0$ parameters, which is interesting, given that this expectation comes from a current generation experiment. The best restriction is again given by the combination with DECIGO, reflected especially in the decreased uncertainties for $H_0$, to $\sigma(H_0) = 0.30$, actually being the best projected constraint on the Hubble constant for all analyses performed.

Now, for the JBP model, we find results that are, in many ways, similar to those obtained for the BA model. The combinations with aLIGO, ET, and DECIGO lead to notable improvements in the cosmological parameters $\Omega_m$ and $H_0$, resulting in a mild reduction in uncertainties relative to the current constraints. In fact, we see no visible improvement in the $w_0$–$w_a$ plane compared to the current constraints. Thus, we notice that for all models, the factor of improvement increases following the sequence of adding aLIGO, ET, and DECIGO to the analysis with the Base data, respectively. While the restrictions on the $w_0$–$w_a$ plane are not significant for this specific model, the analysis with DECIGO leads to good prospects for this form, as we obtain $\sigma(\Omega_m) = 0.0045$ and $\sigma(H_0) = 0.43$, comparable with the results for the other two models.

In general, we note how the choice of parameterization is relevant in our results, reflecting the importance of considering different forms of $w(a)$ in the analyses, as they might represent different classes of well-motivated models that can be compared when using real GWSS data. While models such as JBP are shown to be not so affected by the inclusion of gravitational wave observations, there is a significant contrast when one considers the CPL model, for instance. From the projected confidence contours shown in Fig.~\ref{fig:3}, it is noticeable that, in particular, the improvement given by the mock data in the models is due to the parameter degeneracy breaking present when one usually combines CMB and GWSS data for the $\Lambda$CDM and $w$CDM models. A brief discussion of the impact on the $w$CDM model is shown in Appendix~\ref{appendix:C}. Such degeneracy-breaking power is reduced in $w_0$–$w_a$ models; on the other hand, in the near future, one may expect a more powerful constraint with the inclusion of next-generation CMB data, projected to be available at around the same time as the third-generation gravitational wave collaborations are also obtaining their first results.

\section{Discussion and Conclusions}
\label{sec:6}

Recent advancements in the BAO measurements from DESI~\cite{DESI:2024mwx,DESI:2024lzq,DESI:2025zgx}, when combined with the CMB data from Planck 2018 and three different samples of SNe (PantheonPlus, DESY5, Union3), have indicated evidence for an evolving dark energy component that, at present, behaves like quintessence, but is also compatible with a phantom crossing of $w_{\rm de}(a)$ at higher redshifts. This evidence is primarily based on the assumption of the CPL parameterization of the DE equation of state, but it is also confirmed for other forms of $w_{\rm de}(a)$~\cite{Giare:2024gpk,DESI:2025fii}. Such results raise fundamental questions about the nature of dark energy, as a phantom crossing behavior is not allowed in standard quintessence models, suggesting that alternative scenarios may be necessary to explain the observations.

It is also important to quantify the level of constraint provided by the current DESI data. Although the main results presented by the DESI collaboration emphasize the evidence for DDE, the improvement in the determination of cosmological parameters is also evident. Further combined analyses with current and upcoming datasets will be essential to assess both the reduction in parameter uncertainties and the statistical significance of the evidence for DDE, as demonstrated in the DESI team's studies with different SNe samples~\cite{DESI:2024mwx,DESI:2025zgx}.

In this article, we have examined the improvement in the determination of cosmological parameters when future gravitational wave data are considered in combination with DESI. We have included projections from several major upcoming experiments, such as the ET and DECIGO, as well as results from the current aLIGO. By analyzing three well-known dark energy parameterizations, we have estimated the impact of GW data when added to current constraints including DESI, and we have provided a comparison between the performance of different GW observatories for each model. An interesting outcome of our analysis is that the quality of the constraints is model-dependent. For instance, while ET generally leads to only modest improvements in the $w_0-w_a$ plane, the addition of DECIGO data results in more significant, model-specific gains. In particular, although for models such as the JBP one, the projections were shown to be not very significant in the $w_0-w_a$ plane, the combination of such data, could potentially exclude the $\Lambda$CDM model at several standard deviations, depending on the true underlying cosmology. This highlights the importance of considering a variety of cosmological models when evaluating the constraining power of future datasets.

In conclusion, as expected, Base+DECIGO gives the best results among all parameterizations due to the decreased uncertainties in the sample and the wider redshift range when compared to the aLIGO and ET samples: this is seen especially for quantities such as the matter fraction and the Hubble constant, whose uncertainties are at a similar level to the future CMB projections expected for the next decade, although for $w_0$ and $w_a$ this effect is less pronounced, but present. This is an important issue, as the current DESI evidence for dynamical dark energy models considers the possibility that the equation-of-state parameter was phantom in the not so distant past~\cite{DESI:2025fii}, transitioning to a `quintessence' state at a redshift dependent on the choice of parameterization. This notion has been challenged in some recent works~\cite{deSouza:2025rhv,Dinda:2025iaq}, such that one might consider the universe as consistent with a quintessence-type evolution. This changes the parameter space and points to the necessity of including new data sets that might solve these discrepancies. Our current results support the expectation that future GW data will significantly improve our ability to constrain cosmological parameters. Whether such measurements will further support the standard $\Lambda$CDM paradigm or provide robust evidence for dynamical dark energy remains an open question. In either case, upcoming GW observations will serve as a powerful and independent cross-check of other cosmological probes, particularly those provided by DESI.

\acknowledgments
The authors thank the referee for some important comments that improved the overall quality of the manuscript. FBMS is supported by Conselho Nacional de Desenvolvimento Científico e Tecnológico (CNPq) grant No. 151554/2024-2. JM is supported by Coordena\c{c}\~ao de Aperfei\c{c}oamento de Pessoal de N\'ivel Superior (CAPES). SP has been supported by the Department of Science and Technology (DST), Govt. of India under the Scheme ``Fund for Improvement of S\&T Infrastructure (FIST)'' [File No. SR/FST/MS-I/2019/41]. WY has been supported by the National Natural Science Foundation of China under Grant Nos. 12547110 and 12175096.  EDV is supported by a Royal Society Dorothy Hodgkin Research Fellowship. We acknowledge the use of \texttt{MontePython}, \texttt{CLASS} and \texttt{GetDist}. This work was developed thanks to the use of the National Observatory Data Center (CPDON).
This article is based upon work from COST Action CA21136 Addressing observational tensions in cosmology with systematics and fundamental physics (CosmoVerse) supported by COST (European Cooperation in Science and Technology).

\bibliography{references}

\appendix

\section{Antenna pattern functions for the experiments}\label{appendix:A}

In this appendix, we specify the antenna pattern functions used for each experiment in our modeling of the GW signal detection. We begin with aLIGO, a second-generation gravitational wave detector featuring perpendicular arms arranged in an L-shaped configuration. It is primarily designed to operate within the frequency band of $10\,\mathrm{Hz}$ to $1000\,\mathrm{Hz}$. The antenna pattern functions, which describe the detector’s directional sensitivity to incoming gravitational waves, are given by~\cite{Yang:2021qge}
\begin{align}
    F^{(1)}_{+} &= \tfrac{1}{2}(1 + \cos^2 \theta)\cos 2\phi \cos 2\psi - \cos \theta \sin 2\phi \sin 2\psi, \\
    F^{(1)}_{\times} &= \tfrac{1}{2}(1 + \cos^2 \theta)\cos 2\phi \sin 2\psi + \cos \theta \sin 2\phi \cos 2\psi.
\end{align}

Next, we turn our attention to the ET, a proposed third-generation ground-based gravitational wave detector. Its design features a triangular configuration composed of three interferometers, each with 10 km-long arms, arranged at $60^\circ$ angles relative to one another. The ET is expected to operate over a broad frequency band ranging from $1\,\mathrm{Hz}$ to $10^4\,\mathrm{Hz}$. The corresponding antenna pattern functions for ET are given by~\cite{Califano:2022cmo}
\begin{align}
F_{+}^{(1)}(\theta, \phi, \psi) &= \frac{\sqrt{3}}{2} \left[ \frac{1}{2} (1 + \cos^2 \theta) \cos(2\phi) \cos(2\psi) - \cos\theta \sin(2\phi) \sin(2\psi) \right], \\
F_{\times}^{(1)}(\theta, \phi, \psi) &= \frac{\sqrt{3}}{2} \left[ \frac{1}{2} (1 + \cos^2 \theta) \cos(2\phi) \sin(2\psi) + \cos\theta \sin(2\phi) \cos(2\psi) \right].
\end{align}

The antenna pattern functions for the remaining two antennas are obtained by rotating the azimuthal angle $\phi$ by $2\pi/3$ and $4\pi/3$, respectively:
\[
F_{+,\times}^{(2)}(\theta, \phi, \psi) = F_{+,\times}^{(1)}(\theta, \phi + 2\pi/3, \psi), \quad
F_{+,\times}^{(3)}(\theta, \phi, \psi) = F_{+,\times}^{(1)}(\theta, \phi + 4\pi/3, \psi),
\]
reflecting the $60^\circ$ rotational symmetry of the triangular configuration.

Finally, the third detector considered in this work is DECIGO, a planned Japanese space mission designed to observe gravitational waves in the frequency range of approximately $0.1$–$10$ Hz. The instrument consists of four triangular units. For each unit, in order to obtain orthogonal data streams, linear combinations of the interferometric signals are constructed such that they form an orthogonal basis equivalent to L-shaped interferometers projected on the detector plane. For further details on this configuration, see~\cite{Yagi:2011wg}. The antenna pattern functions for DECIGO are given by
\begin{align}
F^{(1)}_{+}(\theta, \phi, \psi) &= \frac{1}{2} (1 + \cos^2(\theta)) \cos(2\phi) \cos(2\psi) - \cos(\theta) \sin(2\phi) \sin(2\psi), \nonumber \\
F^{(1)}_{\times}(\theta, \phi, \psi) &= \frac{1}{2} (1 + \cos^2(\theta)) \cos(2\phi) \sin(2\psi) + \cos(\theta) \sin(2\phi) \cos(2\psi),
\end{align}
and another pair is $F^{(2)}_{+, \times}(\theta, \phi, \psi) = F^{(1)}_{+, \times}(\theta, \phi - \pi/4, \psi)$, due to the fact that the triangular unit can be effectively regarded as two L-shaped interferometers, aligned at an angle of $45^\circ$ with respect to each other.

For all three detectors, we assume flat distributions for the angular parameters~\cite{Cai:2017aea, Califano:2022cmo}, with $\phi,\psi \in [0,2\pi]$ and $\theta \in [0,\pi]$.

\section{Uncertainties of the generated measurements}\label{appendix:B}

\begin{figure}[t]
    \centering
    \includegraphics[width=0.49\linewidth]{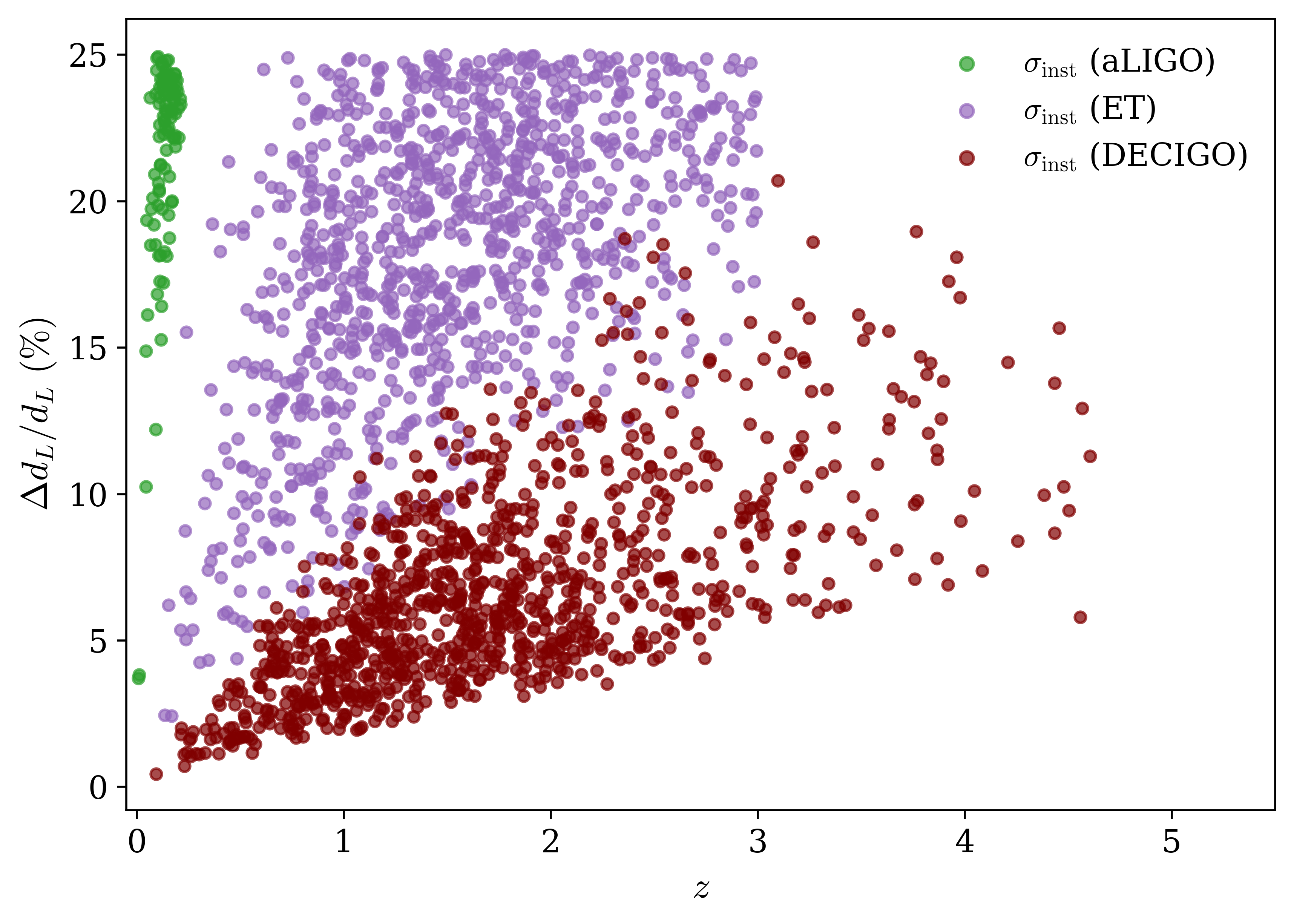}
    \includegraphics[width=0.49\linewidth]{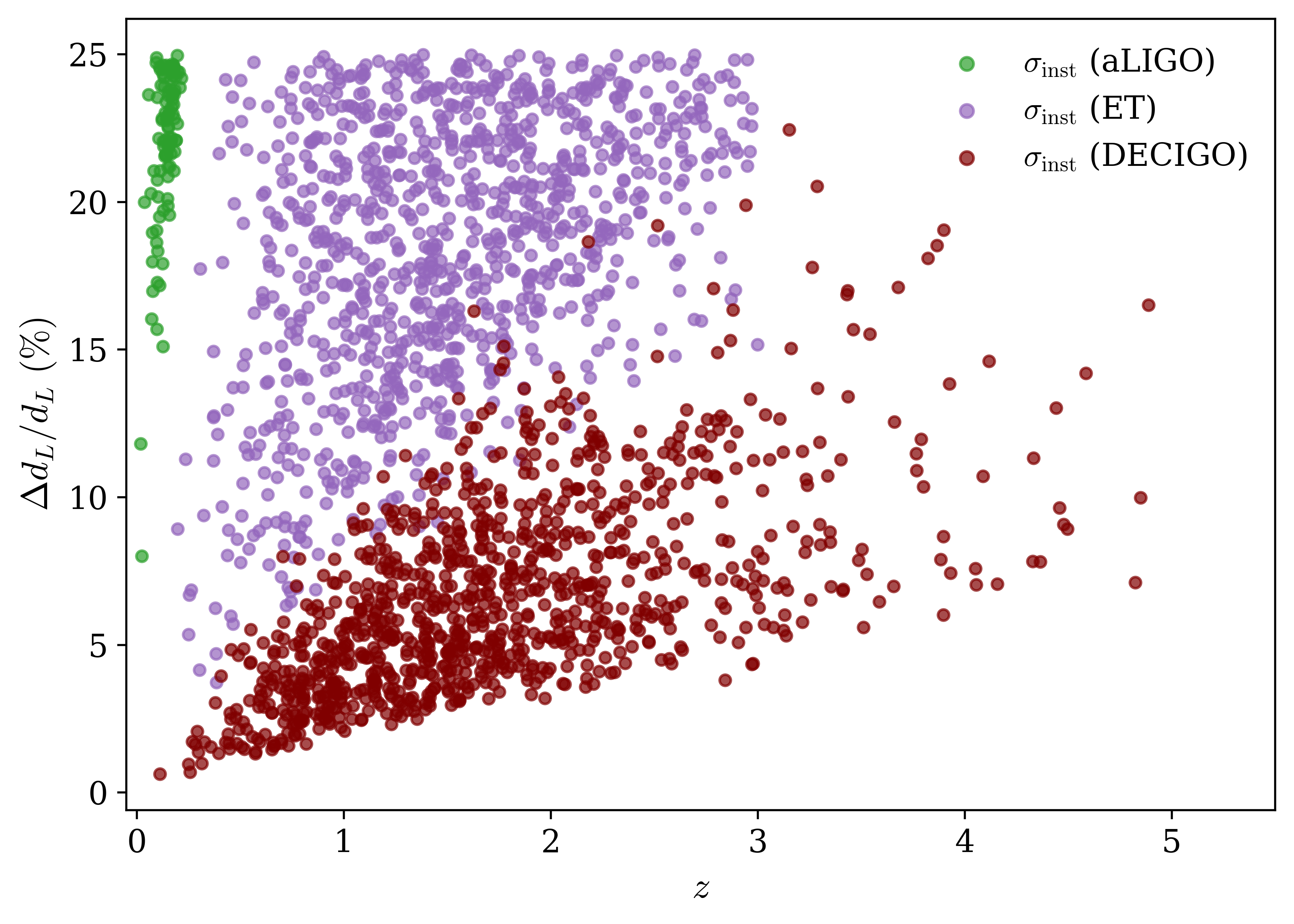}

    \includegraphics[width=0.49\linewidth]{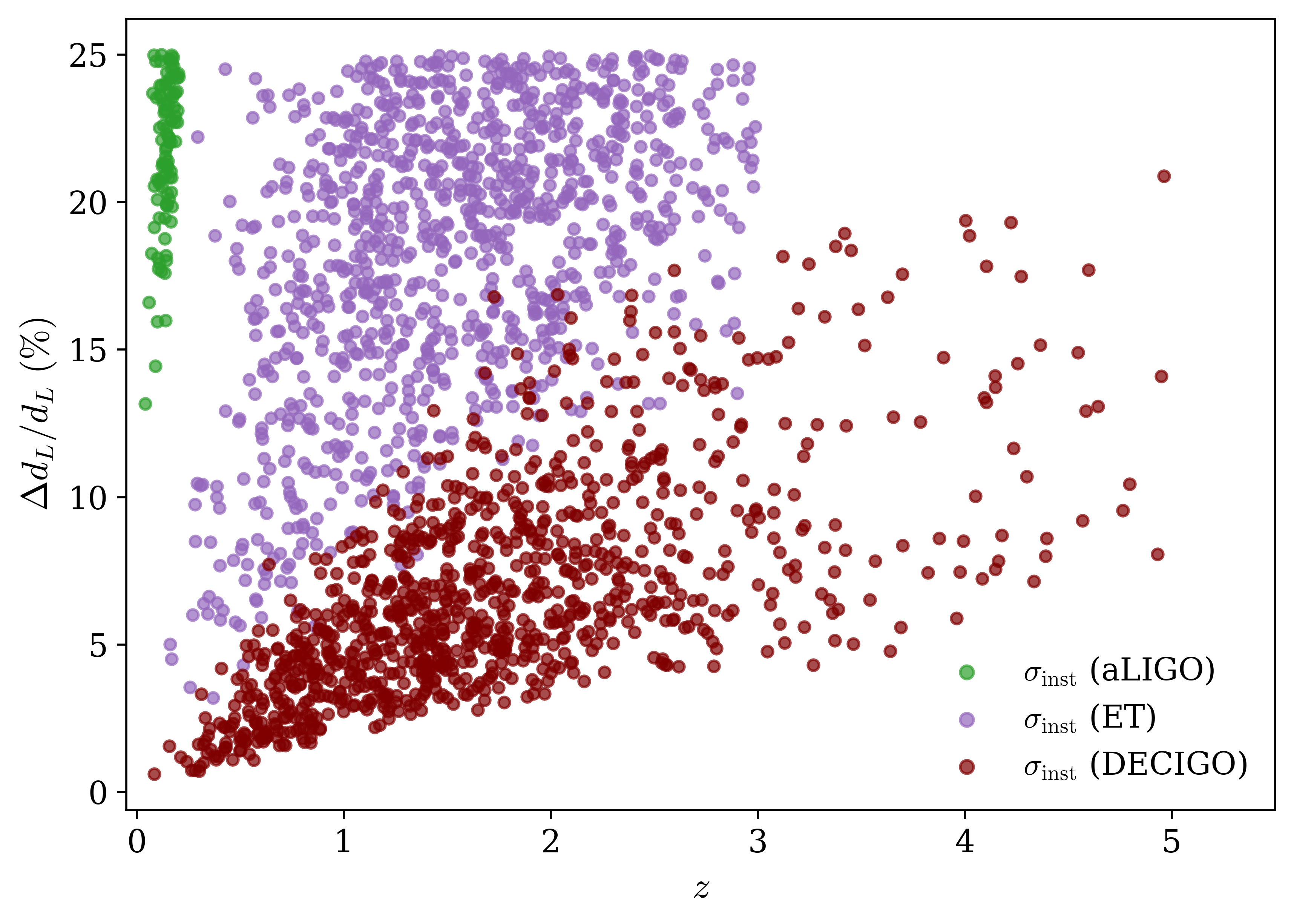}
    \includegraphics[width=0.49\linewidth]{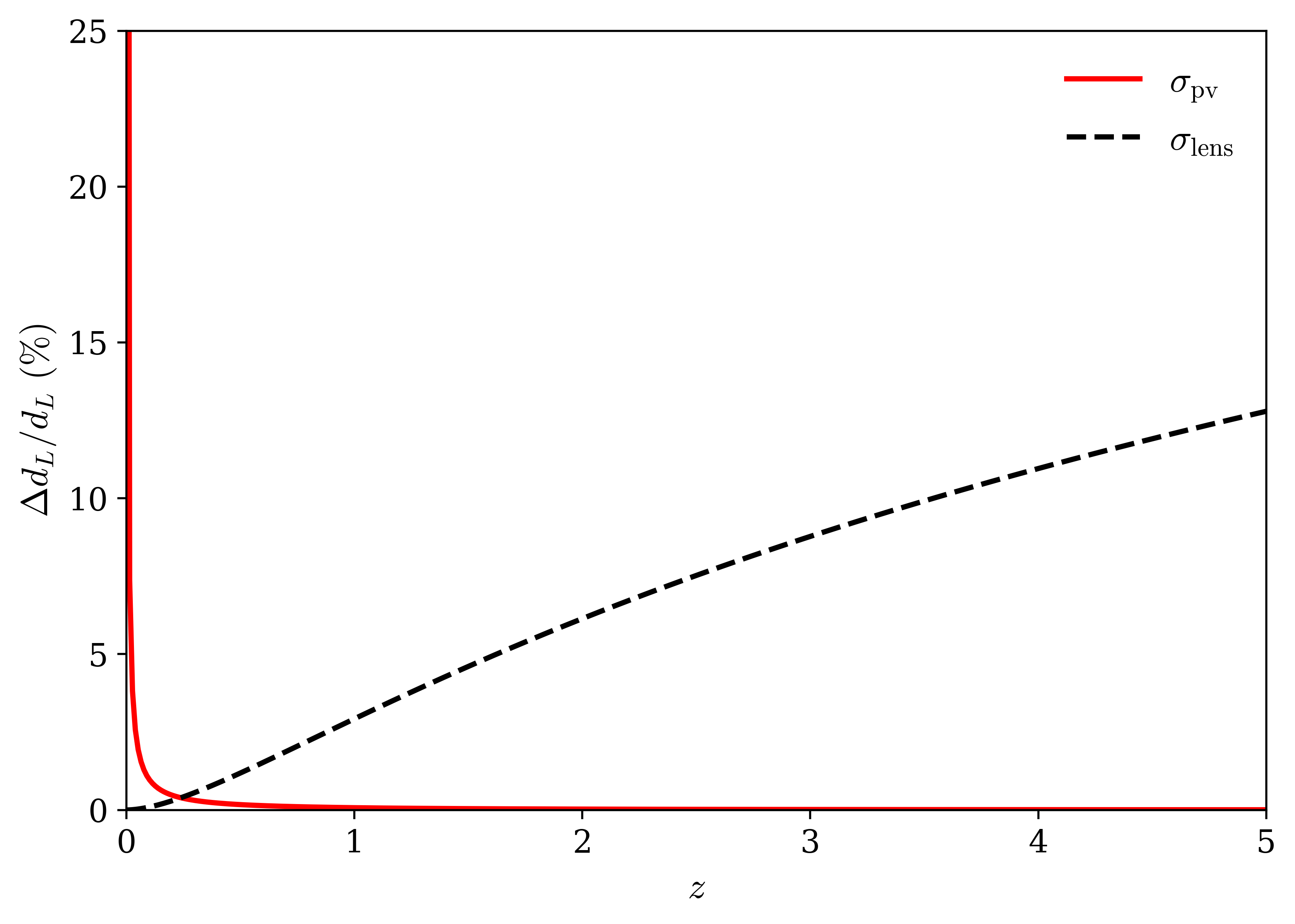}

    \caption{Relative contribution to the uncertainty in the luminosity distance as a function of redshift, expressed as the fractional error $\Delta d_L / d_L$ (in percent), for different scenarios. The first, second, and third panels show the contribution arising solely from instrumental uncertainties for the CPL, BA, and JBP dark energy parameterizations, respectively. The fourth panel (lower right) presents the contribution from peculiar velocity uncertainties and gravitational lensing effects.}

    \label{fig:errors_gws}
\end{figure}

In this appendix, we analyze how each component of the total uncertainty in the luminosity distance, contributes to its measurement. The Fig.~\ref{fig:errors_gws} illustrate the role of instrumental errors, lensing effects, and peculiar velocities of galaxies.

The first, second, and third panels of Fig.~\ref{fig:errors_gws} show the instrumental contribution to the relative uncertainty in the luminosity distance, $\Delta d_L / d_L$, for the CPL, BA, and JBP dark energy parameterizations, respectively. For all parameterizations, aLIGO exhibits the largest instrumental uncertainty, reaching values close to $25\%$ at low redshifts. This is followed by the Einstein Telescope, while DECIGO presents the smallest contribution among the detectors considered. This hierarchy directly reflects the sensitivity of each instrument. Since the instrumental uncertainty can be approximately described as inversely proportional to the signal-to-noise ratio~\cite{Zhang:2019loq}, detectors with lower sensitivity yield smaller SNR values and consequently larger distance uncertainties. As expected, the instrumental uncertainty increases with redshift, as the signal amplitude decreases with luminosity distance, leading to a progressive degradation of the SNR.

The fourth panel of Fig.~\ref{fig:errors_gws} shows the contribution from peculiar velocity and gravitational lensing uncertainties to the relative error in the luminosity distance. The contribution from the peculiar velocity uncertainty, $\sigma_{\rm pv}$, is only relevant at very low redshifts, typically for $z \lesssim 0.01$. This behavior arises because, at low redshifts, the observed redshift is significantly affected by the Doppler motion of the source with respect to the Hubble flow, making peculiar velocities a non-negligible source of uncertainty in the inferred luminosity distance. As the redshift increases, the cosmological expansion dominates over local motions, rendering the peculiar velocity contribution negligible. In contrast, the contribution from the gravitational lensing uncertainty, $\sigma_{\rm lens}$, increases progressively with redshift. This is due to the cumulative effect of large-scale structure along the line of sight.

\section{Impact of mock GWSS data in the $w$CDM model}\label{appendix:C}

\begin{figure}[t]
    \centering
    \includegraphics[width=0.49\linewidth]{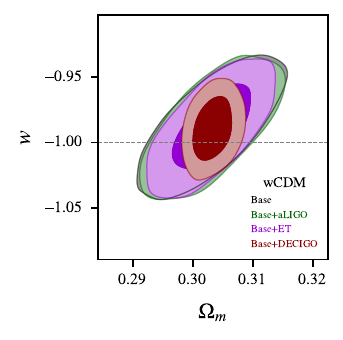}  
    \includegraphics[width=0.49\linewidth]{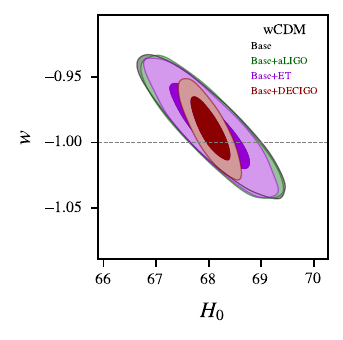}
    \caption{The $\Omega_m-w$ (left) and $H_0-w$ (right) 68\% and 95\% confidence level contours for the $w$CDM model, for all data sets investigated, where we follow the same colors as in Figs.  \ref{fig:2} and \ref{fig:3}.}
    \label{fig:6}
\end{figure}

In what follows, we briefly discuss, for completeness, estimates for the $w$CDM model. The CMB+DESI+PantheonPlus analysis leads to agreement with the $w = -1$ case that reproduces the $\Lambda$CDM model at 68\% C.L., as shown in Fig.~\ref{fig:6}. The impact of adding GW data to the current estimates is clearly shown. When aLIGO is included, the improvement in the uncertainties on $\Omega_m$ is significant, decreasing from $\sigma(\Omega_m) = 0.0051$ to $\sigma(\Omega_m) = 0.0049$, as seen in Table~\ref{tab:3}. The reduction in the errors of the $w$ parameter is not seen, however. When we include projections from ET and DECIGO, however, there is a change of picture, in which the breaking of degeneracies between parameters allows better estimates of both $\Omega_m$ and $H_0$.

\begin{table*}[ht]
		\centering
		\begin{tabular}{>{\scriptsize}c >{\scriptsize}c>{\scriptsize}c>{\scriptsize}c}
			\hline
			\hline
    
            Dataset & $\sigma(\Omega_m)$ & $\sigma(H_0)$ & $\sigma(w)$  \\
            \hline
            \hline 
            \multicolumn{4}{c}{\scriptsize{{$w$CDM}}}\\
            \hline 
            Base & $0.0051$ & $0.57$ & $0.023$  \\
            Base+aLIGO & $0.0049$ & $0.55$ & $0.023$ \\
            Base+ET & $^{+0.0046}_{-0.0041}$ & $0.51$ & $0.022$ \\
            Base+DECIGO & $0.0022$ & $0.25$ & $0.016$ \\
            
            \hline
            \hline
            \hline
		\end{tabular}
		\caption{Same as Table \ref{tab:2} for the $w$CDM model.}
        \label{tab:3}
\end{table*}

In particular, the inclusion of ET and DECIGO decreases the uncertainties of $w$ from $\sigma(w) = 0.023$ to $\sigma(w) = 0.022$ and $\sigma(w) = 0.016$, respectively, with DECIGO providing the best results—an improvement of around 30\% in the uncertainty for the parameter with respect to the current estimate.

\end{document}